\documentclass[useAMS,usenatbib,usegraphicx]{mn2e}
\title[Brown-dwarf secondary stars in CVs]{On the evidence for
brown-dwarf secondary stars in cataclysmic variables}
\author[S. P. Littlefair et al.]{S. P. Littlefair$^{1}$\thanks{E-mail:
sl@astro.ex.ac.uk}, V. S. Dhillon$^{2}$, E. L. Mart\'{\i}n$^{3}$\\
$^{1}$School of Physics, University of Exeter, Stocker Road, 
Exeter EX4 4QL, UK\\
$^{2}$Department of Physics and Astronomy, University of Sheffield, 
Sheffield S3 7RH, UK\\
$^{3}$Institute for Astronomy, University of Hawai`i, 2680 Woodlawn 
Drive, Honolulu, HI 96822, USA}
\begin{document}

\date{Accepted 2002 July 10. Received 2002 July 10; in original form 2002 July 10}

\pagerange{\pageref{firstpage}--\pageref{lastpage}} \pubyear{2002}

\maketitle

\label{firstpage}

\begin{abstract}
We present the K-band spectrum of the cataclysmic variable LL~And,
obtained using {\sc nirspec} on Keck-II. The spectrum shows no
evidence for the absorption features observed by \citet{howell01},
which these authors used to claim a detection of a brown-dwarf
secondary star in LL~And. In light of our new data, we review the
evidence for brown-dwarf secondary stars in this and other
cataclysmic variables.
\end{abstract}


\section{Introduction}

Cataclysmic variables (CVs) are semi-detached binary stars consisting
of a white dwarf primary and a Roche-lobe filling secondary star.
Evolutionary models predict that as the secondary transfers mass to
the white dwarf, the period of the binary star decreases. Eventually,
the mass of the secondary drops below the hydrogen-burning limit and
the secondary star becomes degenerate. This change in the structure of
the secondary star means that further mass loss is accompanied by an
increase in the orbital period \cite[see][ for example]{kolb93}. This
is often used to explain the orbital period minimum which is observed
in CVs at around 80 minutes.

Although for the purposes of this paper we will refer to the secondary
stars in post period-minimum CVs as brown dwarfs, it must be stressed
that they are actually a completely new class of stellar object -- the
degenerate remains of hydrogen-burning stars which have been stripped
of their outer envelopes. \citet{howell97} predict that $\sim70$ per
cent of CVs should have passed the orbital period minimum and contain
brown dwarf secondary stars and yet there is direct spectroscopic evidence
for the existence of only one such object -- LL~And \citep{howell01}. 

LL~And is an extremely faint ($V \sim 20$, $K \sim 18$) dwarf nova.
Its orbital period of 79.2 minutes places it very close to the orbital
period minimum. In this paper we present a higher-quality spectrum of
LL~And which shows no sign of a brown-dwarf secondary star. In light
of this, we review the evidence for the existence of brown-dwarf
secondary stars in CVs.

\section{Observations and data reduction}

On the night of 2001 November 2 we obtained 2.0560--2.4730~$\mu$m
($\sim270$~km\,s$^{-1}$ resolution) spectra of the CV LL~And with {\sc
  nirspec} \citep{mclean98} on the 10-m Keck-II telescope on Mauna
Kea, Hawaii. A total exposure time of one hour was obtained between
airmasses 1.007--1.056 in photometric conditions. The seeing was
approximately 0.7 arcseconds and the slit width was set to 0.76
arcseconds. Observations of the A0V star HD~6457 were also taken to
correct for the effects of telluric absorption and to provide flux
calibration.

{\sc nirspec} introduces curvature and distortion in both the spatial
and dispersion directions. Prior to extraction of spectra, these
effects were removed using the {\sc wmkonspec} package in {\sc iraf}.
Following the removal of the distortion, the nodded frames were
subtracted, the residual sky removed by subtracting a polynomial fit,
and the spectra extracted. There were two stages to the calibration of
the extracted spectra. The first was the calibration of the wavelength
scale using argon arc-lamp exposures; the fourth-order polynomial fits
to the arc lines yielded an error of less than 0.4 angstroms (rms).
The second step was the removal of telluric features and flux
calibration. This was performed by dividing the spectra to be
calibrated by the spectrum of the A0V standard, with its prominent
stellar features interpolated across. We then multiplied the result by
the known flux of the standard at each wavelength, determined using a
black body function set to the same effective temperature and flux as
the standard. As well as providing flux calibrated spectra, this
procedure also removed telluric absorption features from the object
spectra. A final, average spectrum was produced by co-adding the spectra
in the rest frame of the binary centre-of-mass.

\section{Results}

\begin{figure}
\centering
\includegraphics[width=6cm,angle=-90]{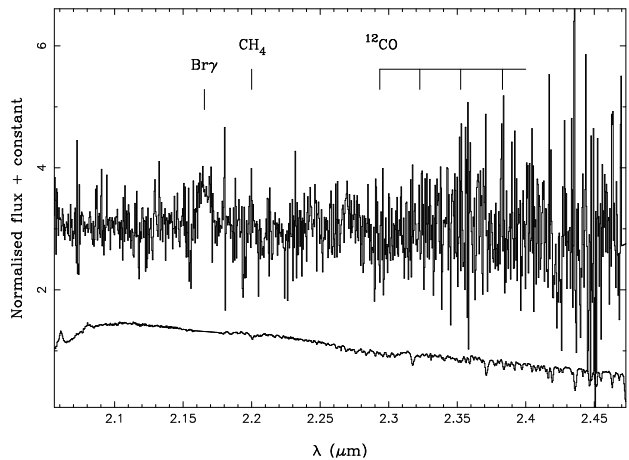}
\caption{K-band spectra of the CV LL~And (top) and the AOV star
HD~6457 (bottom). The spectra have been normalized by dividing by a
first-order polynomial fit to their continua and a $y$-axis offset of
2.0 was then added to the spectrum of LL~And.  The signal-to-noise per
pixel in the LL~And spectrum is $\sim6$. Marked on the plot is the wavelength of the Brackett-$\gamma$ line. Also shown are the wavelengths of the methane and CO features which \protect\cite{howell01} claim to detect.}
\label{fig:figure1}
\end{figure}

Figure~\ref{fig:figure1} shows the Keck spectrum of LL~And in the
K-band. We also show the spectrum of the A0V star HD~6457, which
indicates the location of telluric absorption features. The spectrum
of LL~And is dominated by a strong, broad, single-peaked emission line
of Brackett-$\gamma$ (Br\,$\gamma$), most probably originating in the
accretion disc. The full-width, half-maximum (FWHM) of the line is
$1200\pm200$~km\,s$^{-1}$, compared to the average Br\,$\gamma$ FWHM
in dwarf novae below the period gap of 1500~km\,s$^{-1}$
\citep{dhillon00}. The equivalent width (EW) of Br\,$\gamma$ in LL~And
is $68\pm8$\AA, compared to the average Br\,$\gamma$ EW in dwarf novae
below the period gap of 84\AA\ \citep{dhillon00}. The spectrum of
LL~And is therefore typical of dwarf novae below the period gap.

We see no sign of the secondary star in our spectrum. The spectra is
characterised by a flat continuum and a high level of noise. There is
no marked change in the continuum slope around 2.3 $\mu$m, which would
have been indicative of the presence of water absorption from a
late-type secondary \cite[see][ for example]{dhillon00}. Furthermore,
there is no evidence for the methane and CO absorption bands which
\citet{howell01} claim to have observed in their spectrum of LL~And
(obtained with the {\sc CGS4} spectrograph on the 3.8-m UKIRT
telescope on Mauna Kea).

By convolving the spectrum of LL~And in Fig.~\ref{fig:figure1} with a
K-band filter, we obtain a K-band magnitude of 17.9, very similar to
the value of K=17.5--18 reported in \citet{howell01}. Note that both
of these values only represent lower limits, because they take no
account of slit losses.

\section{Discussion}

\subsection{Is the secondary star in LL~And a brown dwarf?}

On the basis of their proposed detection of methane, water and CO
bands in LL~And, \citet{howell01} claim that this system harbours a
brown-dwarf secondary star. The spectrum of LL~And we present in
Fig.~\ref{fig:figure1} appears to contradict these claims, as it shows
no sign of these features. There are two possible explanations
for this discrepancy:

\begin{enumerate}
  
\item The accretion state of LL And may have changed between the two
observations (which were taken approximately one year apart).
Specifically, our data showing the prominent emission-line from the
accretion disc may have been obtained during a period of high
mass-transfer. If this were the case, the secondary star absorption
features might be swamped by the shot noise from the additional
light. This is unlikely to be true, however, as our K-band magnitude
is almost identical to that of \citet{howell01}, indicating that the
object was in a similar state during the two observations.

\item The signal-to-noise of the \citet{howell01} spectrum may be
significantly worse than in our spectrum. Unfortunately, it is not
possible to estimate the signal-to-noise per pixel in the spectrum of
LL~And presented by \citet{howell01}, because it appears to have been
smoothed. It is possible, however, to predict the relative quality of
the Keck+{\sc nirspec} and UKIRT+{\sc cgs4} spectra using on-line data
for the performance of the two instruments. We find that the spectrum
of \cite{howell01} should have approximately half the signal-to-noise
of our data, which is to be expected given the relative collecting
areas of the telescopes. In order to simulate the effect of this
degradation in signal-to-noise, we added noise to our spectrum so that
the signal-to-noise ratio was halved. We found that this was
sufficient to mask the prominent Br\,$\gamma$ feature in our Keck
spectrum of LL~And and it therefore seems likely that the absorption
features observed by \citet{howell01} are due to noise.
\end{enumerate}
  
In summary,  \citet{howell01} would not have been able
to obtain the signal-to-noise to detect the features claimed. In
addition, for \citet{howell01} to have detected a brown-dwarf
secondary in LL~And, the distance to the CV must be only 30 pc
\citep{howell01}. This is in serious conflict with the lower limit of
364 pc found by \citet{szkody00} and the distance of 760 pc determined
from ultraviolet spectroscopy of the white dwarf by \citet{howell02}.
Hence we believe that there is as yet no direct spectroscopic evidence
for a brown-dwarf secondary star in LL~And.

\subsection{Evidence for brown-dwarf secondary stars in the literature}

So what evidence is there for brown-dwarf secondary stars in CVs? 
In table~\ref{tab:table1}, we list all of the known candidates
from the literature. There are fifteen such systems and the evidence
for their candidature varies in nature and quality, as described below.
\begin{table}
\caption{Brown-dwarf candidates in CVs from the literature. The
central column indicates the methods used to detect the brown dwarf:
RV1 and RV2 refer to the use of the radial velocity of the white dwarf
and secondary star, respectively; SUP refers to the use of the
superhump period-mass ratio relation of \citet{patterson01}; SED
refers to the use of spectral-energy distribution fitting.}
\begin{tabular}{lll}
\hline
& & \\
Object & Evidence & Reference \\
& & \\
\hline
& & \\
1RXS J105010.3-140431 & RV1           & 1       \\
LL And                & lines?        & 2,3     \\
VY Aqr                & SED+lines+SUP & 4,5,6   \\
OY Car                & SUP           & 4       \\
V436 Cen              & SUP           & 4       \\
WX Cet                & SED+SUP       & 4,7     \\
EG Cnc                & SUP           & 4       \\
AL Com                & RV1+SUP       & 4,8     \\
EF Eri                & SED+lines     & 2,9     \\
V592 Her              & SED           & 10      \\
MM Hya                & SUP           & 4       \\
DI UMa                & SUP           & 4       \\
SW UMa                & SED           & 7       \\
WZ Sge                & RV2+SED+SUP   & 4,11,12 \\
HV Vir                & SED+SUP       & 6,7     \\
& & \\
\hline
\end{tabular}
\small

1.~\citet{mennickent01}; 2.~\cite{howell01}; 3.~this paper;
4.~\citet{patterson01}; 5.~\citet{littlefair00};  6.~\citet{mennickent02};
7.~\citet{mason01}; 8.~\citet{howell98}; 9.~\citet{beuermann00};
10.~\citet{vanteeseling99}; 11.~\citet{steeghs01}; 12.~\citet{ciardi98}.
\label{tab:table1}
\end{table}

\subsubsection{Superhump period excess-mass ratio relation}
\label{sec:sup}
The evidence for brown-dwarf secondary stars in many CVs comes from
the relationship between mass ratio and superhump period excess
\citep{patterson01}. Together with an assumed (or measured) value for
the primary mass, this relationship allows one to calculate the mass
of the secondary star for any system with known superhump and orbital
periods.  For systems in which the white dwarf mass was unknown,
\citet{patterson01} assumed the mean white dwarf mass for CVs below
the period gap of 0.7\,M$_{\odot}$.  The calibration of the relationship
between superhump period excess and mass ratio is uncertain, relying
heavily on the assumed mass ratio for WZ Sge. \citet{patterson01}
finds 10 candidates using this method, which are listed in
table~\ref{tab:table1}. Despite the uncertainties in the method, DI
UMa, AL Com, WZ Sge, and EG Cnc should be considered good
candidates, having implied secondary masses below 0.045\,M$_{\odot}$.
The evidence is less compelling in VY Aqr, HV Vir, WX Cet, OY Car,
V436 Cen and MM Hya, in which the implied secondary star masses range from
0.06 to 0.08\,M$_{\odot}$.

\subsubsection{Radial Velocities}
\label{sec:rv}
Three systems have evidence for a brown-dwarf secondary star from
radial velocity studies. In the case of AL Com the evidence is poor;
the upper limit on the white dwarf radial velocity of 32\,km\,s$^{-1}$
places a strict upper limit on the secondary mass of 0.18\,M$_{\odot}$
\citep{howell98}. For 1RXS J105010.3-140431, \citet{mennickent01} find
a white dwarf radial velocity of $4\pm1$\,km\,s$^{-1}$. They claim
that such a low radial velocity for the white dwarf suggests a very
low-mass secondary star (around 10--20 Jovian masses). 
However, the inclination and white dwarf mass are unknown for this
system; a combination of a heavy white dwarf (1.4\,M$_{\odot}$) and
low inclination (5$^{\circ}$) would allow a secondary star mass of up
to 0.11\,M$_{\odot}$.

For the third system, WZ~Sge, \citet{steeghs01} have measured the
radial velocity of the secondary star directly, using
irradiation-induced emission features observed during superoutburst.
Without a reliable estimate of the white dwarf radial velocity, the
authors conservatively derive $M_2 < 0.11M_{\odot}$. Observations to
determine the white dwarf radial velocity in WZ~Sge are hence sorely
needed.

\subsubsection{SED modelling and absorption lines}
\label{sec:sed}
The data used to model spectral-energy distributions in CVs varies
widely from simultaneous, low-resolution spectrophotometry
\citep{mennickent02} and non-simultaneous spectra uncorrected for slit
losses \citep{ciardi98}. Even when using the best data, uncertainties
in deconvolving the contributions from secondary star, white dwarf and
accretion disc make the process highly uncertain.  Therefore, we do
not consider any of the CVs in which the SED provides the only
evidence for a brown-dwarf secondary as good candidates.

In two candidates (VY Aqr \& EF Eri), the absorption lines from the
secondary have also been observed. \citet{mennickent02} find a
spectral type for the secondary star in VY~Aqr of M9.5V, whilst
\citet{howell01} find a spectral type for the secondary star in EF~Eri
of either M6V or L4--5.  It is not possible to say whether these
objects are brown dwarfs solely on the basis of their spectral type as
temperatures may depend on irradiation and the secondary star's
thermal history. It would be possible, however, to obtain a dynamical
mass estimate for these systems by measuring the radial velocities of
the absorption lines. Such a mass estimate would allows us to
determine if the secondary star has been eroded beyond the hydrogen
burning limit.

In summary, there is evidence for a brown dwarf secondary in fifteen
systems. However, there is as yet no {\em direct} evidence (i.e. a
reliable measurement of the secondary star mass) for such an object.
The best candidates are those systems in which an indirect estimate
of the secondary star mass exists (DI UMa, AL Com \& WZ Sge).

\subsection{Infrared photometry of CVs}
\label{sec:2mass}
\begin{figure*}
  \centering \includegraphics[width=13cm,angle=-90]{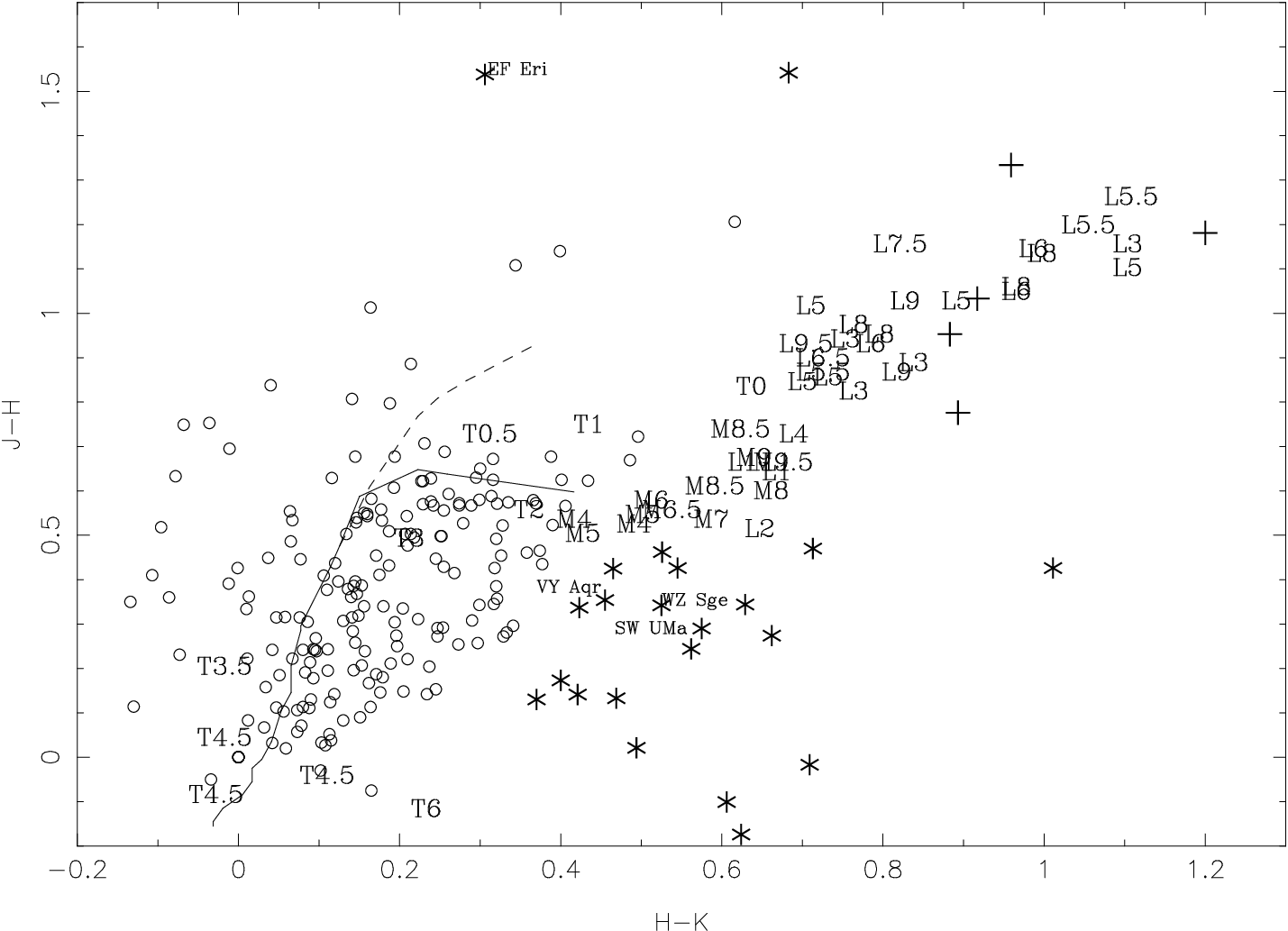}
\caption{Infrared colour-colour diagram for CVs and late-type
dwarfs. The open circles show the CVs from the 2nd incremental data
release of the 2MASS survey \citep{hoard02}. The asterisks show the
CVs which we believe are possible brown-dwarf candidates (see text for
details). Brown-dwarf candidates which we believe might allow a direct
detection of the secondary star are marked with crosses (see text for
details). The positions of late-type dwarfs (from \citealt{leggett02})
in the colour-colour diagram are represented by a text string
indicating their spectral type.  The solid curve shows the position of
the main sequence from spectral-types O9 to M5; the dashed curve shows
the position of the giant sequence from spectral-types G8 to M5
\citep{allens00}. The colours of the main sequence and late-type stars
were put on the 2MASS photometric system using the transformations of
\citet{carpenter01}.}
\label{fig:figure2}
\end{figure*}

\citet{hoard02} recently published infrared photometry for all CVs
contained within the 2MASS 2nd Incremental data release. That data is
plotted in figure~\ref{fig:figure2}. The infrared colours of most CVs
(marked by circles) are easily understood. The secondary star in these
systems is very similar to a late-type, main sequence star. There is also
a blue component of highly variable strength from the accretion flow and
white dwarf. This results in most CVs occupying a region offset bluewards 
(i.e down and left) from near the end of the main sequence.

The 2MASS data release contains photometry for four of the brown-dwarf
candidates in table~\ref{tab:table1}; these systems are labelled in
figure~\ref{fig:figure2}. Three of these four (VY~Aqr, WZ~Sge \& SW
UMa) are significantly redder in $H-K$ than the majority of CVs. The
locations of VY~Aqr, WZ~Sge \& SW UMa in the colour-colour diagram are
consistent with very late-type secondary stars ($\sim$\,M6--L4) and a
blue accretion disc component.  The colours of EF~Eri are remarkable.
Whilst the spectrum of \citet{howell01} shows that the secondary star
dominates the infrared light, EF~Eri is significantly redder in $J-H$
than most CVs. Also, a non-detection of the secondary star in the
optical was used to constrain its spectral type to be later than M9V
\citep{beuermann00}.  Given this evidence, it seems likely that EF Eri
contains a highly unusual secondary star. Other possible explanations
for the unusual infrared colours are that EF Eri is strongly reddened,
or that it may be a triple system.

It can be seen in figure~\ref{fig:figure2} that there is a population
of CVs which occupies the same region as the brown-dwarf candidates
from table~\ref{tab:table1}. This population is distinct from the
population of CVs as a whole. On the basis of their infrared colours,
and bearing in mind the difficulty of deconvolving contributions from
the white dwarf, accretion disc and secondary star to the SED, we
identify these systems as new (weak) brown-dwarf candidates. Our new
candidates were selected according to the colour selection, $J-H \le
3(H-K)-0.8$. They are plotted with asterisks in
figure~\ref{fig:figure2}.  In light of the position of EF~Eri in the
colour-colour diagram, we also identify systems with $J-H > 1.5$ as
new brown-dwarf candidates.

Of the CVs with 2MASS photometry, we identify approximately 10 per
cent as brown-dwarf candidates on the basis of their infrared colours.
This in in contrast to the 70 per cent of systems which are expected
to have passed the period minimum, and contain brown-dwarf secondary
stars.  The small percentage of brown dwarf candidates in the 2MASS
data could be due to selection effects (these systems may be fainter as a
group than most CVs), alternatively, the evolutionary scenario
described in the introduction could be wrong. An alternative scheme
has been suggested in which CVs are too young for there to be many
CVs with brown-dwarf secondaries \citep{pinsonneault02}

\subsection{Future detection of brown-dwarf secondary stars: the prospects}
\label{sec:propsects}
\begin{table}
\caption{Brown-dwarf candidates in CVs from our 2MASS selection, and
those from table~\ref{tab:table1}, ranked according to distance from
the Earth and apparent V-band magnitude (see text for details).}
\begin{tabular}{llccc}
\hline
& & & &\\
Object & \multicolumn{1}{c}{Distance} & 
\multicolumn{1}{c}{Quiescent mag} & 
\multicolumn{1}{c}{$J-H$} &
\multicolumn{1}{c}{$H-K$} \\
& \multicolumn{1}{c}{(pc)} & \multicolumn{1}{c}{(V-band)} \\
& & & &\\
\hline
& & & &\\
VY Aqr                & \hspace*{0.3cm} 34        & 17.1  & 0.354 & 0.455 \\
WZ Sge                & \hspace*{0.3cm} 48$^1$    & 14.5  & 0.342 & 0.525 \\
WX Cet                & \hspace*{0.3cm} 70        & 17.5  & n/a   & n/a   \\
SW UMa                & \hspace*{0.3cm} 110       & 16.5  & 0.290 & 0.575 \\
EF Eri                & \hspace*{0.3cm} $<128^2$  & 18.0  & 1.538 & 0.306 \\
HV Vir                & \hspace*{0.3cm} 149$^3$   & 19.0  & n/a   & n/a   \\
AL Com                & \hspace*{0.3cm} 250$^4$   & 22.0  & n/a   & n/a   \\
OY Car                & \hspace*{0.3cm} 260       & 15.3  & n/a   & n/a   \\
V436 Cen              & \hspace*{0.3cm} 260       & 15.3  & n/a   & n/a   \\
EG Cnc                & \hspace*{0.3cm} 343$^3$   & 19.0  & n/a   & n/a   \\
MM Hya                & \hspace*{0.3cm} 580       & 18.7  & n/a   & n/a   \\
V630 Cyg              & \hspace*{0.3cm} 590       & 17.2  & 0.143 & n/a   \\
LL And                & \hspace*{0.3cm} 760$^5$   & 19.9  & n/a   & n/a   \\
V4140 Sgr             & \hspace*{0.3cm} 1100      & 17.5  & n/a   & n/a   \\
DI UMa                & \hspace*{0.3cm} 1320      & 18.0  & n/a   & n/a   \\

& & & &\\
\hline
\end{tabular}
\small

1.~\citet{spruit98}; 2.~\citet{beuermann00}; 3.~\citet{szkody00}; 
4.~\citet{sproats96}; 5.~\citet{howell02}.
\label{tab:table2}
\end{table}
Combining systems with existing evidence in the literature
(table~\ref{tab:table1}) and systems selected on the basis of their
infrared colours, we now have a list of thirty-nine CVs which may
possess brown-dwarf secondary stars. What are the chances of detecting
the secondary stars in these candidates? Ideally, for a CV to have a
detectable brown-dwarf secondary it should be nearby, and have little
or no ongoing accretion (i.e. it should be optically faint).  With
this in mind, table~\ref{tab:table2} shows distances and V magnitudes
for the thirty-nine brown dwarf candidates. Where no distance was
available in the literature, distances have been estimated using the
relationship which exists for dwarf novae between the absolute
magnitude at outbursts maximum and orbital period. Errors in distances
to SU UMa stars derived from this relationship arise from two sources:
scatter within the relationship and the variations in the maximum
magnitude at outburst for a single system. This can introduce scatter
of up to 1 mag, leading to an uncertainty in the distance of a factor
of $\sim 1.5$ at worst.  If no distance could be obtained, the system
is not listed. $J-H$ and $H-K$ colours are shown, where available to
aid in the location of the objects in figure~\ref{fig:figure2}.

The prospect of direct detection of the secondary star in most of the
candidates shown in figure~\ref{fig:figure2} is remote; $M_K$ for a
spectral type of L3 is $\sim 11$ \citep{leggett02}, whereas the
limiting magnitude (3$\sigma$, 30 minute exposure) of Keck+{\sc
  NIRSPEC} is 18.6. This suggests that a secondary star of spectral
type L3 would only be observable out to distances of 330 pc, even if
there were no contaminating light from the accretion disc or white
dwarf. In CVs, this contamination makes the task of detecting the
secondary much more difficult.  As an example, one of the best
candidates in table~\ref{tab:table2} is WZ~Sge. Extensive infrared
spectroscopy has failed to reveal the secondary in this system
\citep{littlefair00,dhillon00,mennickent02}. It seems likely that in
the majority of systems containing brown-dwarf secondaries, the
secondary stars are too faint, the discs too bright, or the system too
far away for a direct spectroscopic detection in the near-infrared.

There are, however, a few systems in figure~\ref{fig:figure2} which
offer brighter prospects. A number of systems are co-incident with the
colours of late-M and L-type dwarfs, suggesting that a very late-type
secondary dominates the near-infrared light in these systems. These
systems are: Psc3, V529 Ori, 1RXS J114247.5+215717, RX J0502.8+1624 \&
Her\footnote{The system Her has no unique identifier. It is found at
$\alpha = $17 48 0.5.9, $\delta = $34 04 01}. These systems are shown in
figure~\ref{fig:figure2} with a cross.  Of the systems, Psc3, 1RXS
J114247.5+215717 and Her have uncertain designations as CVs
\citep{ritter98}. Also, EF~Eri and another system, V732 Sqr show
highly unusual infrared colours. Follow-up near-infrared spectroscopy
of all these systems is highly desirable.

\section{Conclusions}

\begin{enumerate}
\item We have presented the Keck K-band spectrum of LL~And, and find
  no evidence for a brown-dwarf secondary star.
\item We find no direct evidence for brown-dwarf secondary stars in
  CVs in the literature.
\item Significant indirect evidence exists for brown-dwarf secondary
  stars in DI UMa, AL Com, EG Cnc (from the superhump period excess),
  WZ~Sge (from the superhump period excess and radial velocity of
  the secondary star), and in EF Eri and VY Aqr (from the spectral 
  features of the secondary star).
\item The distances to these CVs imply that the short-term prospects
  of detecting the secondary star is poor. Even in the nearby
  candidate WZ~Sge, infrared spectroscopy has failed to detect the
  secondary star, because of the contribution from the accretion flow
  and white dwarf.
\item There are a small number of CVs whose infrared colours suggest
  that they may contain detectable, brown-dwarf secondary stars.
  However, the prospects of studying large numbers of brown-dwarf
  secondary stars in CVs are bleak.
\end{enumerate}

\section*{Acknowledgments}

We would like to thank Tom Marsh for pointing out the usefulness of
the absolute outburst magnitude-orbital period relationship for
estimating distances to CVs.  We would also like to thank Paul Hirst
and Tim Naylor for invaluable discussions regarding UKIRT pointing.
Data presented herein were obtained at the W. M. Keck Observatory,
which is operated as a scientific partnership among the California
Institute of Technology, the University of California and the National
Aeronautics and Space Administration. The Observatory was made
possible by the generous financial support of the W. M. Keck
Foundation. It is a pleasure to acknowledge the hard work and
dedication of the NIRSPEC instrument team at UCLA: Maryanne Anglionto,
Odvar Bendiksen, George Brims, Leah Buchholz, John Canfield, Kim Chim,
Jonah Hare, Fred Lacayanga, Samuel B. Larson, Tim Liu, Nick Magnone,
Gunnar Skulason, Michael Spencer, Jason Weiss, and Woon Wong. In
addition, we thank the observing assistants at Keck observatory: Joel
Aycock, Gary Puniwai, Charles Sorenson, Ron Quick, and Wayne Wack.

\bibliographystyle{mn2e}
\bibliography{abbrev,refs}

\label{lastpage}

\end{document}